# Revision of nucleotide substitution rate in mtDNA control region of white sturgeon *Acipenser transmontanus* (Acipenseridae)

Shedko S.V.

*Federal Scientific Center of the East Asia Terrestrial Biodiversity,*
*Far Eastern Branch of the Russian Academy of Sciences, Vladivostok, 690022 Russia*
*e-mail: shedko@biosoil.ru*

SUMMARY

The raw data from study of variation of D-loop mtDNA of white sturgeon *Acipenser transmontanus* (Mol. Biol. Evol. 1993. 10: 326–341) was re-analyzed. Re-calculated nucleotide substitution rate (μ) was $0.782$–$0.939 \times 10^{-7}$ substitutions/site/year/lineage, which was 1.4 times less than the estimate given in above-mentioned publication. The use of new μ has led to an increase in estimates of long-term female effective population size ($N_{ef}$) and coalescence times for mtDNA haplotypes, previously calculated for samples of Amur sturgeon *A. schrenckii* and kaluga *A. dauricus*, but it did not affect the conclusions on the critical state of their natural populations.

# Ревизия скорости нуклеотидных замещений в контролирующем регионе мтДНК белого осетра *Acipenser transmontanus* (Acipenseridae)

## С. В. Шедько

*Федеральный научный центр Биоразнообразия наземной биоты Восточной Азии Дальневосточного отделения Российской академии наук, Владивосток, 690022*
*e-mail: shedko@biosoil.ru*

АННОТАЦИЯ

Заново проанализированы первичные данные из работы по анализу изменчивости фрагмента D-петли мтДНК у белого осетра *Acipenser transmontanus* (Mol. Biol. Evol. 1993. 10: 326–341). Рассчитанная по ним скорость нуклеотидных замещений ($\mu$) составила $0.782–0.939 \times 10^{-7}$ замен на позицию/год/линию, что в 1.4 раза меньше оценки, приведенной в указанной публикации. Использование новой $\mu$ привело к росту оценок долговременной эффективной численности самок ($N_{\text{ef}}$) и времени коалесценции мтДНК-гаплотипов, рассчитанных ранее для выборок амурского осетра *A. schrenckii* и калуги *A. dauricus*, но не повлияло на выводы относительно критического состояния их популяций.

*Ключевые слова*: митохондриальная ДНК, скорость эволюции, калибровка, Acipenser.

Скорость – одна из основных характеристик процесса эволюции. Скорость нуклеотидных замещений в последовательности ДНК ($\mu$) – одна из важнейших статистик в молекулярной филогенетике и популяционной генетике. К примеру, на видовом уровне, имея точечные или интервальные оценки $\mu$, полученные для тех или иных узлов ветвления филогенетического дерева с использованием информации об ископаемых остатках, и принимая ту или иную модель варьирования $\mu$ по его ветвям, можно производить датировку всех остальных ветвлений этого дерева.

На популяционном уровне знание $\mu$ позволяет рассчитывать долговременную эффективную численность популяции ($N_{\text{e}}$), основываясь на полученной эмпирической оценке уровня полиморфизма ДНК в ней ($\theta$) и решая связывающее их уравнение $\theta = 4N_e\mu$ (в случае митохондриальной ДНК $\theta = 2N_{ef}\mu$, где $N_{ef}$ – эффективная численность самок в популяции), или исследовать ее динамику во времени в абсолютных значениях, строя байесовские скайлайны [1]. Это представляется особенно важным в области природоохранной генетики тех групп, которые находятся под угрозой исчезновения, поскольку позволяет достичь лучшего понимания текущего состояния их популяций и, соответственно, перспектив на их восстановление. Одной из таких групп являются осетровые рыбы (Acipenseridae), практически все члены которой внесены в Красный список МСОП в статусе видов, требующих охраны.

В последнее десятилетие было установлено, что для одной и той же группы организмов скорость нуклеотидных замещений в ДНК, рассчитанная для длительных (млн или десятки млн лет) интервалов времени, не совпадает с оценками $\mu$, полученными при анализе коротких периодов (десятки, тысячи, или сотни тысяч лет) дивергенции. Как правило, скорость появления мутаций при рассмотрении родословных выше, чем скорость нуклеотидных замещений, оцененная при расхождении популяций, а последняя намного выше, чем скорость, установленная для давно разошедшихся видов (филогенетическая скорость). Это особенно наглядно демонстрирует анализ данных по быстро эволюционирующей митохондриальной ДНК [2]. Поэтому калибровка скорости эволюции ДНК должна, по возможности, производиться в соответствии с временны́м масштабом решаемых задач.

У осетров нуклеотидное разнообразие ($\pi$) в контролирующем участке мтДНК (D-петле) на порядок (в 10–20 раз) выше нуклеотидного разнообразия, свойственного всей остальной части митохондриального генома [3]. Поэтому D-петля, как высоко информативный маркер, часто используется для решения тех или иных задач популяционной генетики осетров.

Для D-петли мтДНК осетров известна лишь одна попытка калибровки молекулярных часов на коротком временном интервале (в пределах голоцен – современность), которая может применяться на популяционном уровне. Она была осуществлена на данных, полученных при анализе изменчивости D-петли мтДНК у белого осетра *Acipenser transmontanus* из рек Колумбия и Фрейзер [4]. В ее основе лежало предположение о том, что бассейн р. Фрейзер был заселен белым осетром сравнительно недавно, после его высвобождения из под Кордильерского ледникового щита из рефугиума, располагавшегося южнее – в бассейне р. Колумбия. Это позволило рассчитать скорость нуклеотидных замещений в фрагменте D-петли мтДНК белого осетра, анализируя его выборки из популяций рек Фрейзер и Колумбия, разделившихся, по мнению авторов этого исследования, 10–12 тыс. лет назад.

Для белого осетра из рек Колумбия и Фрейзер (размер выборок – $n_X = 11$ и $n_Y = 16$, соответственно) авторы приводят следующие оценки ([4]: p. 335): нуклеотидное разнообразие фрагмента D-петли длиной 462 пн в выборках – $d_X = 0.0075$ и $d_Y = 0.0113$; средняя дивергенция между выборками – $d_{XY} = 0.0120$; чистая, или нетто, дистанция между популяциями (за вычетом различий внутри выборок) [5] – $d_A = d_{XY} - (d_X + d_Y)/2 = 0.0026$. Отсюда скорость нуклеотидных замещений ($\mu = d_A/2T$, где $T$ – время дивергенции, равное 10000–12000 лет в данном случае), по их данным, равна $1.09$–$1.31 \times 10^{-7}$ замещений на позицию/год/линию.

Недавно мы обратили внимание на то, что в другом месте этой работы указывается, что нуклеотидное разнообразие, рассчитанное для суммарной выборки последовательностей D-петли белого осетра, было равно 0.0227. Это в два раза выше оценки $d_{XY}$, что вызывает вопросы, поскольку эти показатели, учитывая небольшие различия в размерах выборок из рек Колумбия и Фрейзер, должны быть примерно одного и того же уровня. Кстати, там же в таблице 4 приведено другое значение $\pi$, не отличающееся от $d_{XY}$ – 0.0120. Более того, в защищенной двумя годами ранее диссертации первого из авторов ([6]: p. 76) для того же самого материала можно найти совсем другие цифры: $d_X = 0.00116$, $d_Y = 0.00390$, а $\pi$ для суммарной выборки – 0.01001.

Поскольку те или иные показатели изменчивости D-петли белого осетра не раз выступали в качестве сравнительных в исследованиях других видов осетров, а оценка скорости эволюции D-петли белого осетра использовалась в качестве известного параметра в уравнении $\theta = 2N_{ef}\mu$ или в байесовском анализе коалесценции (см.: [7–9]), то целью настоящей работы была проверка того, насколько корректно они были рассчитаны.

Для решения этого вопроса указанные в работе [4] последовательности 19 гаплотипов D-петли белого осетра (LO1509–LO1527, LO1529) были загружены из базы данных Genbank (NCBI) и выровнены так, как это было сделано в оригинальном исследовании (рисунок). Далее на основе информации из той же работы были восстановлены их выборочные частоты.

Произведенные на этой матрице данных с помощью программы MEGA6 [10] расчеты привели к следующим результатам (использовалось $p$-расстояние,

позиции с делециями отбрасывались только в попарных сравнениях, ошибка рассчитывалась по результатам 500 циклов бутстрепа): $d_X = 0.01413 \pm 0.00310$, $d_Y = 0.01998 \pm 0.00406$, $d_{XY} = 0.01847 \pm 0.00369$, π для объединенной выборки – $0.01831 \pm 0.00365$, $d_A = 0.00141 \pm 0.00058$ (где x и y – выборки из рек Колумбия и Фрейзер, соответственно).

В итоге полученная нами оценка μ – $0.588–0.705 \times 10^{-7}$ замещений на позицию/год/линию – оказалась практически в два раза ниже той, что была приведена в работе Брауна и соавторов [4]. Таким образом, анализ исходных данных не подтверждает оценок, опубликованных в оригинальном исследовании. В последнем, судя по всему, были допущены грубые ошибки в расчетах параметров изменчивости D-петли у белого осетра. Стоит отметить, что теперь филогенетически близкие (сестринские) виды – белый осетр и амурский осетр *A. schrenckii* – оказываются сходны не только по уровню гаплотипического, но и по уровню нуклеотидного разнообразия (см. [8]).

Использованное Брауном и соавторами [4], а также нами *p*-расстояние не учитывает важных особенностей эволюции последовательностей мтДНК (нуклеотидный состав, соотношение транзиций и трансверсий, неоднородное распределение замен по позициям и т.д.). Поэтому для данного набора данных была подобрана оптимальная модель нуклеотидных замещений – HKY+I, отобранная с помощью программы PAUP v. 4.0a152 [11] из 56 возможных вариантов на основе скорректированного информационного критерия Акаики (AICc).

Оценка μ, рассчитанная на основе модели HKY+I, оказалась несколько больше оценки, полученной нами при использовании *p*-расстояния – $0.782–0.939 \times 10^{-7}$ замещений на позицию/год/линию. Тем не менее, она все также была существенно (примерно в 1.4 раза) меньше оценки, сообщенной Брауном и соавторами [4].

Чтобы провести коррекцию оценок $N_{ef}$, полученных ранее для амурского осетра и калуги [8–9], для выборок последовательностей D-петли мтДНК этих видов осетров были заново построены байесовские скайлайны с использованием новой оценки μ – $0.861 \times 10^{-7}$ (средняя интервала $0.782–0.939 \times 10^{-7}$). При этом была задействована такая же, как и для фрагмента D-петли белого осетра, модель нуклеотидных замещений – HKY+I. Все остальные условия соответствовали тем, что описаны ранее [8–9].

В результате расчетная $N_{ef}$ для популяции амурского осетра составила 36207 особей (95%-й интервал наивысшей апостериорной плотности, HPD: 13643–135076), а для популяции калуги примерно в 9 раз меньше – 3900 особей (95% HPD: 778–20555). Следует подчеркнуть, что так как оценки $N_{ef}$ увеличились в 1.4–1.5 раз, то все сделанные ранее заключения по критическому состоянию популяций амурского осетра и калуги не потеряли своей силы, поскольку контраст между оценками долговременной эффективной численности и текущей численностью самок у этих видов осетров лишь усилился.

Отметим, что в новых условиях (другая μ, модель нуклеотидных замещений HKY+I вместо HKY) оценки времени коалесценции гаплотипов, рассчитанные для выборок последовательностей D-петли у амурского осетра и калуги, выросли в 1.5–1.8 раз. Для амурского осетра этот период составил 286770 (95% HPD: 161320–444190) лет, а для калуги – 41650 (95% HPD: 17010–80830) лет. При этом оцениваемый из байесовских скайлайнов характер изменения кривых численности этих видов во времени остался прежним, и заключение о неодинаковой демографической истории популяций амурского осетра и калуги [9] также сохранило свою силу.

## СПИСОК ЛИТЕРАТУРЫ

```
                              11111222223333334444
                           111222333567006660245634455801 45
                           546924756770827467111155190416601   X   Y
L01510_Cr1_F6              CAA-TTCAAAGGACGTATTAGTTTCGAA-TG-C       1
L01511_Cr2_UC3_F0          .......GG-..GT.C...G.C..T.......T   1   1
L01512_Cr3_F3              .......GG-..GT.C...G.C.CT.......T       1
L01513_Cr4_F3              .......GG-..GT-C...G.C...T......T       1
L01521_Cr12_F1             .......GG-..GT-C...G.C..........T       1
L01523_Cr14_F3             .......GG-..GT.C...G.C..TA......T       1
L01526_Cr17_F1             .......GG-..GT.C...G.C..........T       1
L01520_Cr11_F9             T...C....-AAG...GC.........-.TCA.T      1
L01525_Cr16_F11            T........-AAG...GC......T.G.TCA.T       1
L01524_Cr15_LC10           T......G.-A.G....C......T...T.A.T   1
L01514_Cr5_F7_F2           .G...CT..-....A.....G..CCT......AT      2
L01515_Cr6_2xUC8           .G...CT..-....A.....G.CCCT......T   2
L01516_Cr7_LC2             .G...C...-....A.....G..CCT......AT  1
L01519_Cr10_F2             .G....T..-..........G..CCT......AT  1
L01522_Cr13_F2_2xLC_UC2    .G...CT..-....A.....G..CCT......T   3   1
L01527_Cr18_UC2            .G...CT..-....A...CG..CCT.......T   1
L01517_Cr8_LC4             .G...CT..-....A.....CT..........T   1
L01518_Cr9_3xF4            .GTG.CT..-..........CT..........T       3
L01529_Cr19_UC6            ....................A......G.....   1
```

**Рис.** Вариабельные позиции в фрагменте *D*-петли мтДНК белого осетра, гаплотипы и их встречаемость в выборках из рек Колумбия (X) и Фрейзер (Y).

**Figure**. Variable positions in mitochondrial control region of white sturgeon, haplotypes and their occurrence in samples from Columbia (X) and Fraser (Y) rivers.